\documentstyle[pra,epsf,aps,twocolumn]{revtex}
\def\aut#1{#1}
\begin{document}
\sloppy
\def\bpi{\begin{picture}}
\def\epi{\end{picture}}

\author{H.~Kleinert%\thanks{E-mail kleinert@physik.fu-berlin.de,~\newline
%URL http://www.physik.fu-berlin.de/\~{}kleinert}
                   \\
         Freie Universit\"at Berlin\\
          Institut f\"ur Theoretische Physik\\
          Arnimallee14, D-14195 Berlin
     }
\title{Criterion for Dominance
of
Directional
over Size Fluctuations
in Destroying Order
}
\maketitle
\begin{abstract}
For  systems exhibiting a  second-order phase transition
with a spontaneously broken continuous O($N$)-symmetry at low temperature,
we give a criterion for judging at which temperature  $T_K$
long-range directional fluctuations of the
order field destroy the order when approaching the
critical temperature from below.
The temperature $T_K$ lies always significantly
below the famous Ginzburg temperature $T_G$
at which
size fluctuations of finite range
in the order field become important.
\end{abstract}
%%%%%%

{}~\\
{\bf 1.}
Although the fluctuation behavior of systems
in a second-order phase transition
is universal in the
immediate vicinity of the transition,
and the theory is well established \cite{Kad},
there are
still disputes
concerning the dominant fluctuation
mechanism which drives
a number of important phase transitions.
A typical
example is the question whether
the superfluid transition in liquid helium
is initiated by the proliferation of vortex lines, which
are the
defects
in all systems with pure
angular fluctuations of a complex order parameter,
or by the size fluctuations of the associated complex {\em order field\/}
$\phi(x)$.
At first sight, this question may seem meaningless
since
all  critical exponents
of the transition can be calculated
with great accuracy either from an order field
theory with $\phi^4$-interaction
\cite{Sat}, where size and directional fluctuations
seem to be equally important,
or from
Heisenberg model on a lattice, which contains
only directional fluctuations.
The equivalence can easily be proved for superfluids,
where
the Heisenberg model reduces to an
XY-model.
After a duality transformation, the XY-model
can be reexpressed
as a sum over a grand-canonical ensemble of non-selfbacktracking
vortex lines  \cite{Peskin,GFCM1}
whose proliferation completely describes all properties of the
superfluid  transition.
They produce the same critical exponents
as a
complex $\phi^4$-theory, and the reason for this is simple:
the XY-model may
 be converted
into a complex $\phi^4$-theory
by a transformation of integration variables
in the functional integral
of the partition function  \cite{GFCM1}.

The question whether angular or size fluctuations
drive a phase transition
does therefore not
concern the  immediate vicinity of the transition
where the properties are
universally governed by a
critical exponent $ \omega $.
It is only a meaningful
question
in the {\em precritical\/}
regime of the transition.
where a mean-field description of a system breaks down.
There
it possesses an answer similar to the Ginzburg criterion
which estimates the temperature range where this happens
due to {\em size fluctuations\/} of the order field.
The new criterion
to be presented in this note
will tell us where  the breakdown is caused by
{\em directional fluctuations\/}.
In the case of a complex order field, these
lead to an early proliferation of vortex lines
before size fluctuations become large.

The new criterion serves to understand
the dominance of
directional fluctuations
in the recently-discovered restoration of continuous symmetry
in Gross-Neveu \cite{KB} and Nambu--Jona-Lasinio models \cite{KBo},
and, more importantly, the generation of a pseudogap phase
above the superfluid phase
in
strong-coupling superconductors \cite{BK}.

{}~\\
{\bf 2.}
In order to lead up to the new criterion,
we briefly recall the relevant features of
the field-theoretic approach to
the critical exponents in the immediate vicinity of
second-order phase transitions\cite{Kad}.
		   Since critical properties are
caused by the long-wavelength
fluctuations of a system,
it is sufficient to identify
these, assign to each of them a real order field $\phi_A(x)$,
$A=1,\dots,N$,
the local generalization of Landau's order parameter \cite{Lan},
and set up a Ginzburg-Landau energy
density. Its fluctuations
are
studied
with the help of
a functional integral
over all field configurations,
weighted by a Boltzmann factor of the total field energy.

At a moderate temperature distance $|T-T_c|$ from the critical
temperature $T_c$,
the functional integral may be evaluated
by the saddle point approximation.
This is the mean-field regime of the field theory.
When approaching the
critical point, the fluctuations of the order parameter
increase. They become important in a
temperature regime  $ \Delta T_G$ around $T_c$,
whose width was first  estimated
by the famous Ginzburg criterion \cite{Ginz}:
It is the regime where the fluctuations
of the order parameter
in pockets
of coherence size
 reach into the normal phase.

The  Ginzburg-Landau field energy density
is found from phenomenological studies
of a system
in some neighborhood of the
critical temperature
\cite{Lan}.
One expands the energy density in powers
of the order field and its derivatives.
Then one identifies the temperature interval
where all expansion terms
are irrelevant except for those appearing in a simple
$\phi^4$-theory.
At some {\em mean-field critical temperature\/} $T^{\rm MF}_c\neq T_c$,
correlation lengths become infinite in the mean-field approximation.
The mass
term goes through zero
linearly in $T/T_c^{\rm MF}$, i.e., the bare
square
mass
$m^2$ is proportional to $\tau \equiv T/T^{\rm MF}_c-1$.

As the system enters the critical regime,
fluctuations of the order parameter become important
and must be accounted for.
This is done
by perturbative methods.
Each physical quantity of interest is expanded into a Taylor series
of the reduced interaction strength $g/m^{4-D}$,
and a resummation of these
in the limit $m\rightarrow 0$
% most conveniently
%by the strong-coupling theory  of Ref.~\cite{sc}
allows us to extract
power laws $m^{\rm power}$, which determine the
critical exponents  \cite{Parisi}.

We shall now demonstrate that
in systems with a continuous symmetry,
the Ginzburg criterion is not sufficient
to characterize  the temperature range where fluctuations
are important.
A system will exhibit strong
directional fluctuations before the size fluctuations of the order parameter
become noticeable, leading
to a
phase transition at a temperature $T_K$
far below the
Ginzburg
temperature $T_G$.

{}~\\
{ 3.}
For simplicity, we restrict the argument to
$N$ order fields $\phi_A$ with
O($N$)-symmetry in $D$ dimensions.
With a convenient choice of field and mass normalization,
the
Ginzburg-Landau energy density in $D$ dimensions may be written
as
\begin{eqnarray}
 \varepsilon (\!\phi_A,\!\partial \phi_A\!)\!=\!\frac{1}{2a^D}\!\left\{ \!
\alpha ^2\! a^2
\!\left[\partial \phi_A(x)\right]^2
\!+\!\tau
\phi^2_A(x)
\!+\!\frac{g}{2}
\left[ \phi^2_A(x)\right]^2 \!\right\}
{}.    \nonumber
\label{@ner}
\end{eqnarray}
{}From here on we use natural
units with $k_BT_c^{\rm MF}=1$.
The fields have zero engineering dimension,
$a$
denotes some
microscopic length scale of the system, usually the size of atoms
or molecules,
and $g$ is some interaction strength.
The parameter $ \alpha $
specifies
the zero-temperature coherence length
 of
the system in units of $a$ as being
$\xi_0= \alpha a/ \sqrt{2}$.
This can vary greatly from system to system.
In superconductors, for example, $ \alpha $
it can lie anywhere between a few thousand,
and less than  ten in high-temperature superconductors.

In this note, we shall only be concerned
with the destruction of the ordered state which lies
{\em below\/} the critical temperature
where $\tau <0$: There the fields in the energy
density
fluctuate around an ordered  ground state with
a constant vector
 $\langle\phi_A \rangle\equiv$ $\Phi_A\equiv $ $\langle\phi \rangle \,N_A\equiv
$ $\Phi N_A$
in field space, whose direction vector $N_A$ breaks spontaneously the
O($N$)-symmetry,
and whose magnitude is
$\Phi=\sqrt{\Phi_A^2}= \sqrt{-\tau /g}$,
where the energy density is minimal,
fluctuating around the condensation energy
density
$ \varepsilon _{0}= \varepsilon (\Phi_A,0)=-\tau ^2/4ga^D$.
The temperature-dependent coherence length
%
%\begin{equation}
$\xi=  { \alpha a}/{\sqrt{2|\tau|}}$
%\label{@xi}\end{equation}
%
describes the range of
the size fluctuations of the order field.

The magnitude
is estimated by assuming the
field to live in patches on a simple cubic lattice
of spacing $\xi_l=l\xi$, choosing eventually
a spacing parameter $l=2$ to ensure the independence
of the patches.
Then
\begin{equation}
\langle [\phi(x)\!-\!\Phi]^2\rangle
/ \Phi^2
%=\left(\frac{a}{s\xi}\right)^{D-2}\!\!\!\!\!\frac{v_1(0)}{ \alpha ^2}
=l^{2-D}(2|\tau  |)^{D/2-2} g\alpha ^{-D}v_1(0),
\label{@fluc}\end{equation}
where $v_{m^2}^D(0)=\int _{-\pi}^\pi d^Dk/[ \sum _{i=1}^D(2-2\cos k_i)+m^2]$
is the lattice Coulomb potential
of reduced mass $m$.
For $D=3,4,\dots, ...$, $v_1^D(0)$ has the values \cite{vc}
\begin{eqnarray}
v_1^D(0)\approx 0.1943,~~0.1270,~~~\dots,~~~1/2D.
\label{@}\end{eqnarray}
Mean-field behavior breaks down if (\ref{@fluc})
is of the order unity, which happens at the reduced
Ginzburg temperature
\begin{equation}
  |\tau _G|\approx [K\,v_1^D(0)/l]^{2/(4-D)} ,~~~~~D<4
{}.
\label{@GI}\end{equation}
where
\begin{equation}
K \equiv 2^{D/2-1}g/ \alpha ^D,
\label{@YE}\end{equation}
i.e., at a
 Ginzburg temperature
 $ T_G\equiv T^{\rm MF}_c (1-| \tau _G|)$.
Ginzburg, in his original paper \cite{Ginz},
estimated
$v_1^D(0)$ in three dimensions
by an
integral
  $\int {d^3p}/(2\pi)^3(p^2+ 1)
\approx
  (2\pi^2)^{-1}\int_{0}^\pi dp\,p^2/(p^2+ 1) \approx1/4\pi$,
and assumed $l=1$,
which lead to $|\tau _G|\approx (g/ \alpha ^3)^2/8\pi^2$.
In old-fashioned type-II superconductors,
$|\tau _G|$ can be as small as
 1/10000 \cite{315},
which explains
why conventional superconductors are well described by mean-field
theory.
In modern high-$T_c$ superconductors,  on the other hand,
Ginzburg's estimate leads to $|\tau _G|\approx0.01 $ \cite{Lobb},
such that critical exponents become observable.

For $D>4$, the right-hand side in (\ref{@fluc}) decreases when approaching the
critical point,
so only mean-field behavior is observed.
If
$D=4-  \varepsilon $
 lies only slightly below four,
the right-hand side of (\ref{@GI})
behaves like $|\tau |^{ \varepsilon /2}$,
implying a good mean-field description
until $|\tau |$ is
extremely small.

{}~\\
{ 4.}
The derivation of the new criterion
is based on the observation
that
the
kinetic term
defines a second, completely independent,
  energy scale of the system.
For its identification, we split
the fields according to size and direction in O($N$) field space as
$%\begin{equation}
\phi_A=\phi\, n_A,~n_A^2=1.
$ %\label{@}\end{equation}
The directions $n_A$ describe the long-range
fluctuations of the Goldstone modes.
Sufficiently far from the
critical regime, we
may neglect the gradient term of the size  $\phi(x)$,
and approximate the energy density by
\begin{eqnarray}
\varepsilon (\phi,\partial n_A)\!=\!\frac{1}{2a^D}\left\{  \!\alpha ^2 \!a^2
\phi^2(x)\left[\partial n_A(x)\right]^2
\!+\!\tau
\phi^2(x)
\!+\!\frac{g}{2}
 \phi^4(x)\!
\right\}
\!.
\label{@ner2}      \nonumber
\end{eqnarray}
The fluctuations of the Goldstone modes are controlled
by the gradient term whose magnitude
 depends on the size $\Phi$ of $\phi$ at the minimum
of the potential.
The gradient energy density is
\begin{equation}
 \varepsilon_{n_A}(\partial n_A)=\frac{ \beta }{2\xi_l^{D-2}}[\partial
n_A(x)]^2,
\label{@enst}\end{equation}
with
$ \beta = \beta(\Phi)$=$
\alpha ^2(\xi_l/a)^{D-2}  \Phi^2
$=$
{\alpha ^Dl^{D-2}}/$${ (2|\tau |)^{D/2-2}}{g}.
$
This is the second energy scale. It measures how much energy is
spent
when reversing
the direction vector $n_A$ over the distance $\xi_l$, and is called
the {\em stiffness\/}
of the directional field.

{}From studies of O($N$)-symmetric classical Heisenberg models it is known
that directional fluctuations disorder
a system if the bending stiffness drops below a certain
critical value $ \beta _{\rm cr}$.
For large $N$, this value can easily be estimated
by a simple manipulation of the functional integral
for the partition function
associated with the energy
(\ref{@enst}).
It may be written as \cite{ZJ}
\begin{eqnarray}
Z_{n_A}\!=\!
\int\!\! {\cal D}n_A
 {\cal D} \lambda
e^{-  (\beta /2\xi^{D-2})
\int d^Dx \left\{ [ \partial n_A(x)]^2\!+ \!\lambda (x)
[ n^2_A(x)-\!1]\right\}}. \nonumber
\end{eqnarray}
where
 the unit length of
$n_A(x)$ is enforced by a Lagrange
multiplier field $ \lambda (x)$.  Integrating out the $n_A({ x})$-fields
leaves us with a pure $ \lambda ({ x})$-field theory,
with an energy functional
\begin{equation}
    E[\lambda]  = \frac{N}{2}{\rm Tr}\log[-\partial ^2+ \lambda({ x} )]-
 \frac{ \beta  }{2 \xi^{D-2}} \int d^Dx \, \lambda ({ x}).
\label{@nfl}\end{equation}
For large $N$, the fluctuations of the $ \lambda $-field are
frozen, and the disordered state
has an energy density given by the extremum of (\ref{@nfl}), where $ \lambda ({
x})$
is a constant
satisfying the gap equation
$%\begin{eqnarray}
 \beta = \beta _ \lambda
  \equiv N v^D_ \lambda (0).
$ %\label{@}\end{eqnarray}
The order is destroyed at a critical stiffness
at
$ \beta _{\rm cr}= \beta _0$.
On a simple cubic lattice, we find
in three, four, and large-$D$ dimensions
\cite{vc}:
\begin{eqnarray}
 \beta_{\rm cr}=
{N}v_0^D({ 0}) \approx
N\, 0.2527\,,~~~N\,0.1549,~~~N/2D
,          \label{@stic}
\label{@critst}\end{eqnarray}
respectively.
Formula (\ref{@critst}) is reliable only for large $N$.
However, Monte Carlo simulations show that
the critical value (\ref{@stic}) can be trusted
already for $D=3$ and $N=2$
 to within about 10\%, where
 simulations
yield
$ \beta _{\rm cr}^{\rm MC}\approx
0.45$ (see \cite{montecarl}),  in good agreement with
the value
$0.5054$
from (\ref{@stic}).
The simulations are done
by
putting the Heisenberg model on a lattice of unit spacing,
so that the
energy density
for $N=2$ takes the
XY-model
form
$%\begin{equation}
 \varepsilon _{n_A}(\partial n_A)\approx { \beta }
\sum_{\mu=1,\dots, D}
 [1-\cos\nabla _\mu \gamma (x)],
$ %\label{@XY}\end{equation}
where
 $\nabla_i$ denotes
the lattice gradient in the $i$th coordinate direction, and $ \gamma \equiv
\arctan n_2/n_1$.
Since the quality of the approximation increases with $N$
and $D$, we can trust Eq.~(\ref{@critst})
to within 10\% for all $N$ and $D\ge3$.
This accuracy will be sufficient for the criterion
to be derived here.

\begin{figure}[b]
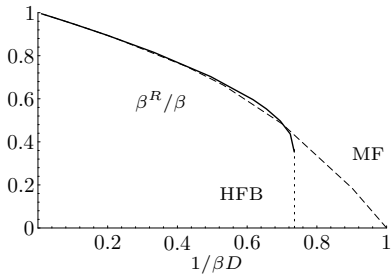

\vspace{-.5cm}~~~~\input pl.tps
\vspace{.4cm}
\caption[]{Softening of stiffness of
XY-model
%according to Eq.~(\protect\ref{@sca}),
derived
from a self-consistent
approximation \`a la Hartree-Fock-Bogoliubov.
%At $ \beta ^{cr}$ the stiffness collapses
%due to a restoration
%of the spontaneously  broken symmetry.
The dashed curve shows the mean-field approximation
given by $ \rho _s= \alpha ^2/4D^2 \beta ^2$,
$ \beta = \alpha I_0( \alpha )/2DI_1( \alpha )$ [$I_n( \alpha )$ = modified
Bessel functions]
which goes to zero linearly in $|\tau|$.
 The exact stiffness goes  to zero like $|T_c-T|^{(D-2) \nu }$,
with the critical exponent $ \nu $
($\approx 0.6705$ for $D=3$ \cite{Sat}).
}
\label{fig1}\end{figure}~\\[-2em]%
The critical stiffness can, incidentally, be also
estimated by calculating  its
renormalized version from a
sum of an infinite number
of terms
in a perturbation expansion.
Expanding the cosine into a Taylor series, and
calculating the harmonic expectation values of
 quartic, sextic, etc. terms,
we find in
a
self-consistent  approximation of the
Hartree-Fock-Bogoliubov type
that the stiffness
has a renormalized value \cite{stiff}
%
%\begin{equation}
 $\beta_R= \beta \,e^{-1/2D \beta_R}.$
%\label{@sca}\end{equation}%
%
This softens with increasing temperature $1/ \beta $, until
$ \beta $ reaches a critical value
$ \beta _{\rm cr}=e/2D$ where $\beta_R$ drops to zero (see Fig.~\ref{fig1}).
In $D=3$ dimensions,
this happens at
$ \beta _{\rm cr}=0.4530\dots~$,
a value which is
in excellent agreement with the Monte Carlo number $ \beta _{\rm cr}^{\rm MC}
\approx
0.45$.
The prediction of such sharp drop is true only in
two dimensions, as shown by Kosterlitz and Thouless \cite{KT}.
For $D>2$ it
 is an artefact
of the approximations,
and the exact $\rho _s$  goes to zero like $|T_c-T |^{(D-2) \nu }$,
with a critical exponent $\nu\approx1/2+(4-D)/10+\dots~$.

The estimate for the critical stiffness
(\ref{@stic}) leads now directly to the announced criterion:
The phase fluctuations
will disorder the system
if the stiffness $\beta$
in Eq.~(\ref{@enst})
drops below the critical
value (\ref{@stic}),
which happens at a reduced temperature
\begin{equation}
  |\tau _K|\approx [NK\,v_0^D(0)/l]^{2/(4-D)} ,~~~~~D<4,~N\ge 2.
\label{@Kli}\end{equation}
Thus we obtain the important result that
\begin{eqnarray}
  ~|\tau _K|\approx [Nv_0(0)/v_1^D(0)]^{2/(4-D)}  |\tau _G|,~~D<4,~N\ge 2.
\label{@Klir}\end{eqnarray}
This implies that for all systems with
$N\ge 2$,
directional fluctuations destroy the order
{\em before\/}  size fluctuations become large.
They cause a phase transition below the Ginzburg temperature
at
 $  T_K\equiv T^{\rm MF}_c (1-| \tau _K|)$.
For $D=3$, the relation becomes $  |\tau _K|\approx 2.184\, N^2  |\tau _G|$.
Thus, if the critical regime is
approached in a $\phi^4$-theory
with a well-formed mean-field regime,
the transition
is {\em always\/} initiated by
directional fluctuations.
In particular, the estimates for the critical regime
of the high-$|T_c|$ superconductors \cite{vc}
will receive  a factor $\approx9$.

The dominance of directional fluctuations
is, of course, most prominent
the limit of large $N$,
and it is therefore not surprising
that the critical exponents of
the $\phi^4$-theory and the Heisenberg model have the same $1/N$-expansions
in any dimension $D>2$,
as a pleasant demonstration of the universality
of critical phenomena.

By adding to the field energy density
$\varepsilon (\phi,\partial n_A)$
 the energy density of directional fluctuations
with the field-dependent
stiffness
$ \beta = \beta (\phi)= \alpha ^D\phi^2l^{D-2}/ (2|\tau|)^{D/2-1} $
we can study, as in Ref.~\cite{KBo},
the combined energy density
in the disordered phase where the symmetry is restored
but the average   $\Phi$ of the size of the order field $\phi$
is nonzero.

{}~\\
{\bf 5}.
How do we determine experimentally
the fluctuation parameter $K$ to estimate $|\tau _G|$ and $|\tau _K|$?
In magnetic systems, one measures the susceptibility tensor
 $\chi_{AB}( { k})\equiv \int d^Dx e^{i{ kx}}\langle
\phi_A(x)\phi_B(0)\rangle$
 at wave vector ${ k}$,
and decomposes it
into
parallel and perpendicular parts
%$\chi_{\parallel}( { k})$ and $\chi_{\perp}( { k})$
as $
\chi_{AB}( { k})=(\Phi_A\Phi_B /\Phi^2) \chi_{\parallel}( { k}  )+
( \delta _{AB}-\Phi_A\Phi_B/\Phi^2) \chi_{\perp}( { k})$.
The mean-field behavior of these quantities are
$\chi_{\parallel}( { k})\approx a^3/( \alpha ^2a^2{ k}^2+2|\tau |) $
and $\chi_{\perp}( { k})\approx a^3/\alpha ^2a^2{ k}^2$.
Combining these at ${ k}=0$ with
the mean-field behavior of the spontaneous magnetization
 $\Phi= \sqrt{|\tau |/g}$, and with the temperature-dependent
coherence length
$\xi$,
we see that the size of $K$ can immediately be
estimated from a plot,
versus $t\equiv T/T_c-1$,
 of either of the
dimensionless
experimental quantities
\begin{eqnarray}
K_{\rm exp}
\approx\left.{|t |^{2-D/2}}\frac{{ k}^2}{\xi^{D-2}}
\frac{\chi_{\perp}({ k})}{k_BT\Phi^2}\right|_{{ k}\rightarrow 0}
\!\!\!\!\!\!{\rm or}\,~{|t |^{2-D/2}}{\xi^{D}}\!\!
\frac{\chi_{\parallel}(0)}{k_BT\Phi^2}
,          \nonumber
\label{@cr2}\end{eqnarray}
these being written down in in physical units.
Note that $t$ measures the temperature distance
from the experimental $T_c$, in contrast to
$\tau \equiv T/T^{\rm MF}_c-1 $.
In the mean-field regime, where $t \approx \tau $, $K_{\rm exp}$ is
constant,
and can be inserted into
Eq.~(\ref{@Kli})
to find
the temperature $T_K$
where
directional fluctuations
destroy the order.

In superfluid helium we may plot, in analogy to the transverse
susceptibility expression for $K_{\rm exp}$, the quantity
$%\begin{equation}
K_{\rm exp}\approx
{|t|^{2-D/2}}
{M^2k_BT}/{\xi^{D-2} \hbar^2\rho _s} ,
$ %\label{@crhe}\end{equation}
where  $M$ is the atomic mass and
$ \rho _s $ the superfluid mass density, which at the mean-field level is
defined by
writing the gradient energy
(\ref{@enst}) as
$ %\begin{equation}
 %\varepsilon_{n_A}(\partial n_A)=
({  \rho _s }/{2k_BT})
({  \hbar^2}/{M^2})
[\partial n_A(x)]^2.
$ %\label{@}\end{equation}
In the critical regime, the three expressions
for $K_{\rm exp}$
go universally to zero like $|t |^{2-D/2}$,
since $\xi\propto |t |^{- \nu },~\chi_{\parallel}({ 0})\approx|t |^{( \eta
-2)
\nu },~
{ k}^2\chi_{\perp}({ k})|_{{ k}\rightarrow 0}\approx|t |^{ \eta  \nu },~
\Phi^2\approx|t |^{ \nu (D-2+ \eta )},
\rho _s\approx|t |^{(D -2)\nu }$, with $ \eta \approx
[(N+2)/2(N+8)^2](4-D)^2+\dots~$.

Experimentally, the superfluid density of
helium for $D=3$ shows no mean-field behavior
a la Ginzburg-Landau down to
$T\approx T_c/4$, such that the above formulas cannot properly be applied.
Let us
nevertheless estimate
orders of magnitude of a would-be
mean-field behavior:
 $ \rho _s/ \rho \approx 2|\tau |$ \cite{428}, where
$ \rho =M/a^3$ is the total mass density, with $a\approx3.59$\,\AA \cite{257}.
Then the  factor $k_BT_c$  at $T_c=2.18$\,K can be expressed as
$k_BT_c\approx2.35 \hbar^2/Ma^3$ \cite{257}.
With $\xi_0\approx 2$\,\AA,
we obtain
an estimate
$K\approx
1.2\,{a}/{\xi_0}\approx 2.$
Inserting this into
Eq.~(\ref{@Kli}) and relation (\ref{@Klir}),
we obtain
\begin{equation}
|\tau _K|\approx 1/4,~~~~
|\tau _G|\approx 0.03.
\label{@}\end{equation}
The large size of $|\tau _K|$ reflects the bad quality
of a mean-field description.

{}~\\
Acknowledgment: \\
The author is grateful to
B. Van den Bossche
for many discussions.
and to  Drs. E. Babaev, J. Jersak,
A. Pelster, A. Schakel, K. Wiese
for useful comments.

\end{document}